# Twisted bi-layer magnetic photonic crystals

You-Ming Liu,[1] Shi-Kai Lin,[1] Pei-Shi Li,[1] Yi-Ran Hao,[1,*] Biao Yang,[1,#]

[1] College of Advanced Interdisciplinary Studies, National University of Defense Technology, Changsha 410073, China

\* *yhaoaf@connect.ust.hk*
\# *yangbiaocam@nudt.edu.cn*

**In photonics, twisted bi-layer systems have demonstrated unprecedented control over light-matter interactions, primarily through the modulation of photonic band structures and the formation of Moiré patterns. Meanwhile, magnetic photonic crystals have served as cornerstone platforms for manipulating light propagation, facilitating key applications such as Faraday rotation-based isolators and non-reciprocal devices. Nevertheless, the synergistic integration of twist engineering and magneto-optical effects in bi-layer architectures remains unexplored. This work introduces twisted magnetic bi-layer photonic crystal slabs as a novel platform to unify these degrees of freedom. By continuously tuning the twist angle between two magneto-active photonic layers, the giant circular dichroism is observed, and the transmitted waves can be perfectly linearly polarized and rotated. These effects arise from the interplay between resonant properties of the Moiré cell and magnetization-dependent coupling of circularly polarized states. This work establishes a foundation for magnetic topological photonics, bridging twistronics and magneto-optics to unlock new mechanisms for dynamic light control in compact and reconfigurable devices.**

The proposal of Moiré superlattices has paved the way for a novel research frontier, demonstrating that the relative rotation between two-dimensional lattices can lead to properties absent in the monolayers [1-4]. Following the pioneering work of Cao et al. on flat bands [5] and superconductivity [6] in twisted bi-layer graphene, more exotic properties have been unveiled, including topological states [7, 8], exciton engineering [9, 10] and thermal management [11-13].

This concept has been extended to photonics through the design of twisted bi-layer photonic crystals (TBPCs). The formation of Moiré superlattices in TBPC results in a larger lattice constant than in monolayer systems, thereby modifying their scattering characteristics. Lou et al. developed a model to predict the resonant chiral behavior of TBPCs [14]. Tang et al. demonstrated a remarkably adjustable band structure in the TBPC [15]. Subsequent research has further advanced practical applications based on TBPCs, such as computational imaging [16] and the filter [17]. Beyond the electromagnetic (EM) properties, TBPCs also exhibit profound topological characteristics. Wang et al. observed reconstructing Fermi arcs by constructing twisted photonic Weyl meta-crystals [18], and Zhang et al. demonstrated the existence of bound states in the continuum within a TBPC system [19].

On the other hand, magneto-optical (MO) photonic crystal slabs represent a highly attractive platform. The MO material breaks the time-reversal symmetry, thereby inducing nonreciprocal propagation [20-23]. This property makes them particularly suitable for designing optical isolators and circulators [24-28]. In photonic band engineering, the external magnetic field generates a topological edge mode in the bandgap, which supports unidirectional wave propagation and suppresses backscattering [29-33].

However, the convergence of these fields in the form of twisted bi-layer magnetic photonic crystals (TBMPCs) remains unexplored. This work aims to fill this gap by systematically investigating the EM properties of TBMPCs. We focus on the circular dichroism of TBMPCs and the Faraday rotation of the zeroth-order diffraction waves at various incommensurate twist angles.

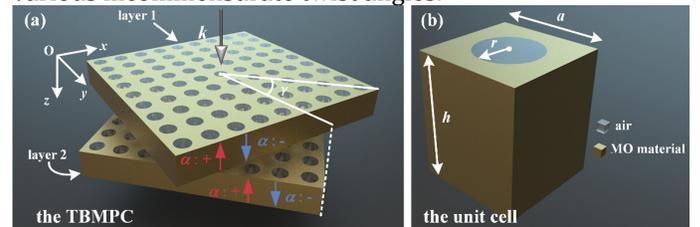

**Fig. 1.** (a) Schematic of TBMPC. An EM wave with wave vector $k$ is normally incident into the system. With layer 1 fixed, the relative twist angle between layer 1 and layer 2 is defined by $\gamma$. The parameters $\alpha$ represents the magnetic field strength (by magnitude) and direction (by sign), respectively. (b) The unit cell of the MO photonic crystal, with $a$=18 mm, $h$=0.25$a$ and $r$=0.25$a$. The following parameters are fixed throughout this study: $\varepsilon=\varepsilon_z=9$, $\mu=\mu_z=1$ and $\beta=0$.

The designed TBMPC is shown in Fig. 1. Each layer consists of a square lattice formed by air cylindrical rods periodically embedded in an MO material slab. The radius of the air rod is $r=0.25a$, where $a$ denotes the lattice constant, equal to 18 mm, and the thickness of the slab $h$ is $0.25a$. The thickness of the air gap inserted between two photonic crystal slabs is also $0.25a$. We define $\gamma$ as the twist angle of layer 2 relative to layer 1. An EM wave is normally incident along the $+z$ direction.

The relative permittivity and permeability can be expressed in the form of tensors [34]:

$$\overleftrightarrow{\epsilon} = \begin{pmatrix} \epsilon & i\alpha & 0 \\ -i\alpha & \epsilon & 0 \\ 0 & 0 & \epsilon_z \end{pmatrix} \text{ and } \overleftrightarrow{\mu} = \begin{pmatrix} \mu & i\beta & 0 \\ -i\beta & \mu & 0 \\ 0 & 0 & \mu_z \end{pmatrix}. \quad (1)$$

In our simulations, the parameters $\alpha$ and $\beta$ represent the applied magnetic field: their magnitudes correspond to the field strengths, and their signs indicate their directions. Unless otherwise specified, the parameter $\alpha$ has the same value in both layers.

We take use of the rigorous coupled wave analysis (RCWA) to investigate TBMPCs [35-37]. The Maxwell equations can be reformulated, consistent with those reported in Refs. 35 and 36:

$$\frac{d}{d\tilde{z}}\begin{bmatrix} \boldsymbol{u_x} \\ \boldsymbol{u_y} \end{bmatrix} = \boldsymbol{Q}\begin{bmatrix} \boldsymbol{s_x} \\ \boldsymbol{s_y} \end{bmatrix} \text{ and } \frac{d}{d\tilde{z}}\begin{bmatrix} \boldsymbol{s_x} \\ \boldsymbol{s_y} \end{bmatrix} = \boldsymbol{P}\begin{bmatrix} \boldsymbol{u_x} \\ \boldsymbol{u_y} \end{bmatrix}. \quad (2)$$

where $\boldsymbol{u_x}$, $\boldsymbol{u_y}$, $\boldsymbol{s_x}$ and $\boldsymbol{s_y}$ are the vectors comprising the diffraction orders of the magnetic and electric fields. Due to the limited computing power, they are both 625-dimensional in this paper.

The only differences are:

$$\boldsymbol{P} = \begin{pmatrix} \widetilde{\mathbf{k}}_\mathbf{x}[\epsilon_z]^{-1}\widetilde{\mathbf{k}}_\mathbf{y} - i[\beta] & -\widetilde{\mathbf{k}}_\mathbf{x}[\epsilon_z]^{-1}\widetilde{\mathbf{k}}_\mathbf{x} + [\mu] \\ \widetilde{\mathbf{k}}_\mathbf{y}[\epsilon_z]^{-1}\widetilde{\mathbf{k}}_\mathbf{y} - [\mu] & -\widetilde{\mathbf{k}}_\mathbf{y}[\epsilon_z]^{-1}\widetilde{\mathbf{k}}_\mathbf{x} - i[\beta] \end{pmatrix}, \quad (3)$$

$$\boldsymbol{Q} = \begin{pmatrix} \widetilde{\mathbf{k}}_\mathbf{x}[\mu_z]^{-1}\widetilde{\mathbf{k}}_\mathbf{y} - i[\alpha] & -\widetilde{\mathbf{k}}_\mathbf{x}[\mu_z]^{-1}\widetilde{\mathbf{k}}_\mathbf{x} + [\epsilon] \\ \widetilde{\mathbf{k}}_\mathbf{y}[\mu_z]^{-1}\widetilde{\mathbf{k}}_\mathbf{y} - [\epsilon] & -\widetilde{\mathbf{k}}_\mathbf{y}[\mu_z]^{-1}\widetilde{\mathbf{k}}_\mathbf{x} - i[\alpha] \end{pmatrix}. \quad (4)$$

Here, $[\epsilon_z]$, $[\epsilon]$ and $[\alpha]$ are the convolution matrices of the permittivity distributed in the space, so are $[\mu_z]$, $[\mu]$ and $[\beta]$. $\widetilde{\mathbf{k}}_\mathbf{x}$ and $\widetilde{\mathbf{k}}_\mathbf{y}$ represent the diagonal matrices containing the wave vectors of diffraction orders. $\tilde{z} = k_0 z$ denotes the scaled propagation distance.

Then the field components are solved as:

$$\begin{bmatrix} \boldsymbol{S_x}(\tilde{z}) \\ \boldsymbol{S_y}(\tilde{z}) \end{bmatrix} = \boldsymbol{W}e^{-\Lambda\tilde{z}}\boldsymbol{c}^+ + \boldsymbol{W}e^{+\Lambda\tilde{z}}\boldsymbol{c}^-. \quad (5)$$

where $\boldsymbol{W}$ and $\boldsymbol{\Lambda}^2$ are the eigenvector matrix and the diagonal eigenvalue matrix of $\boldsymbol{\Omega}^2 = \boldsymbol{PQ}$.

The EM fields between two interfaces (of the i$^{th}$ layer) are associated by the scattering matrix:

$$\begin{bmatrix} \boldsymbol{c}_{in}^- \\ \boldsymbol{c}_{out}^+ \end{bmatrix} = \boldsymbol{S}^{(i)}\begin{bmatrix} \boldsymbol{c}_{in}^+ \\ \boldsymbol{c}_{out}^- \end{bmatrix} = \begin{bmatrix} S_{11}^{(i)} & S_{12}^{(i)} \\ S_{21}^{(i)} & S_{22}^{(i)} \end{bmatrix}\begin{bmatrix} \boldsymbol{c}_{in}^+ \\ \boldsymbol{c}_{out}^- \end{bmatrix}. \quad (6)$$

For TBPC, chaining the matrix of each layer by the Redheffer star operator $\otimes$ [35, 36], the scattering matrix of the device satisfies:

$$\boldsymbol{S} = \boldsymbol{S}^{(1)} \otimes \boldsymbol{S}^{(2)}. \quad (7)$$

The transmission and reflection information can be derived from the scattering matrix.

Fig. 2 shows the transmission spectra for the left-circularly polarized (LCP) and right-circularly polarized (RCP) EM waves. In this work, the LCP are defined by the condition wherein the electric field along the $y$-axis precedes that along the $x$-axis by a phase of $\pi/2$. The RCP EM waves are opposite in rotation.

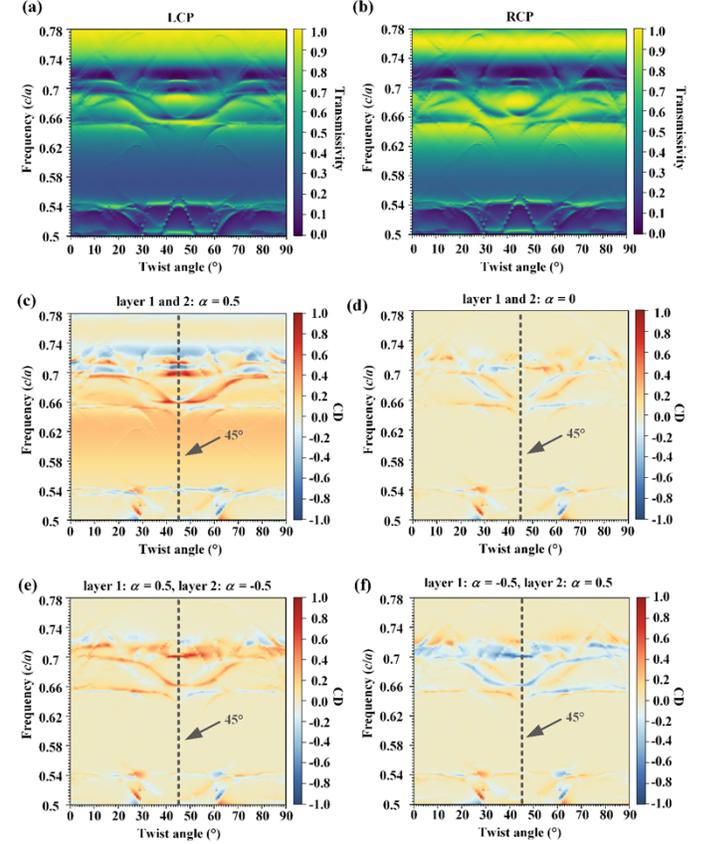

**Fig. 2.** (a, b) Transmissivity as a function of twist angle and frequency for LCP and RCP incident waves, respectively. The parameter $\alpha$ is set as 0.5 in both layers. (c-f) CD = ($T_{RCP}$ - $T_{LCP}$) / ($T_{RCP}$ + $T_{LCP}$) as a function of twist angle and frequency, which is defined to analyze the circular dichroism. The value of $\alpha$ follows such conditions: (c) 0.5 in layer 1 and layer 2; (d) 0 in layer 1 and layer 2; (e) 0.5 in layer 1 and -0.5 in layer 2; (f) -0.5 in layer 1 and 0.5 in layer 2.

Pronounced transmission maxima typically manifests the resonant modes of the photonic structure. In Fig. 2(a), the resonant effects are more prominent in two frequency regions of the studied band, approximately from $0.5c/a$ to $0.55c/a$ and $0.64c/a$ to $0.71c/a$ ($c$ is the light speed). In the high-frequency range above $0.74c/a$, the incident LCP waves can be almost completely transmitted independent of the twist angle. However, the resonant frequencies shift with the twist angle because modifying the angle alters the Moiré unit cell, thereby affecting the scattering characteristics. Considering the introduction of the MO material in TBPC, the effective refractive index is different for the LCP and RCP waves. Thus, the transmission spectra for the RCP are distinct from those of the LCP. In Fig. 2(b), this effect is more significant at higher frequencies.

$T_{RCP}$ and $T_{LCP}$ represent the transmissivity for the RCP and LCP waves. Here, we define CD= ($T_{RCP}$ - $T_{LCP}$) / ($T_{RCP}$ + $T_{LCP}$) to analyze the circular dichroism. Some magnitudes of CD are extremely high at the specific frequencies. It means that only one mode of polarization (LCP or RCP) can be preferentially transmitted, while the other will be significantly blocked. As displayed in Fig. 2(c), notable chiral behavior is observed at the resonance positions. Compared to low frequencies from $0.5c/a$ to $0.55c/a$, high frequencies ranging from $0.64c/a$ to $0.71c/a$ exhibit better circular dichroism over a broader range of twist angles. The twist angle also reveals the tuning effects on the circular dichroism.

At 0° or 45° twist angles, the system possesses the $C_{4v}$ symmetry. As shown in Fig. 2(d), for a general TBPC, the spectral lines are expected to exhibit odd symmetry with respect to 0° or 45° [14]. However, the different relative permittivity of the MO material for LCP and RCP waves disrupts this symmetric characteristic as shown in Fig. 2(c). The CD maps in Figs. 2(e) and (f) correspond to conditions with swapped $\alpha$ values between layer 1 and layer 2. Comparing Figs. 2(c) and (e), a reversal of the magnetic direction in layer 2 is accompanied by a reduction in CD across most of the area, while regions of significant CD become distinctly highlighted.

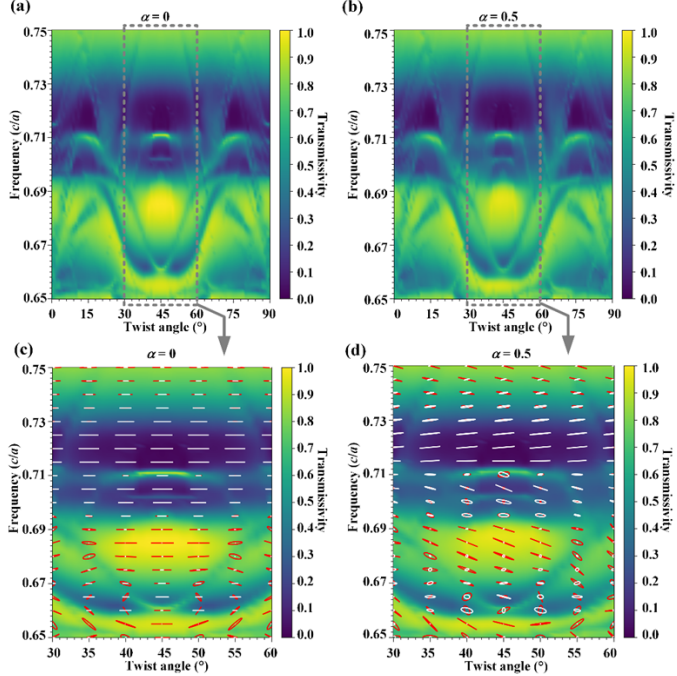

**Fig. 3.** Evolution of the transmissivity with the twist angle (0° to 90°) for (a) $\alpha = 0$ and (b) $\alpha = 0.5$. Polarization states of the zeroth-order transmitted (red) and the reflected (white) EM waves for (c) $\alpha = 0$ and (d) $\alpha = 0.5$. The polarization state is observed along the wave propagation direction.

Fig. 3 reveals the Faraday rotation of the zeroth-order diffraction waves with the incident EM waves polarized along the $x$-axis. In Fig. 3(a), the transmission spectra are symmetric about 45°, as are the polarization states in Fig. 3(c). Comparing Figs. 3(a) and (b), the transmissivity remains almost unchanged after applying an external magnetic field. While in Fig. 3(d), the magnetic field rotates the transmitted and reflected polarization states. Linearly polarized transmission is observed in high-transmissivity regions, whereas the polarization becomes elliptical in low-transmissivity regions. In convention, the Faraday rotation is usually analyzed by $\theta = V \cdot B \cdot L$, where $\theta$, $B$, $L$ and $V$ are the rotation angle of polarization, the magnetic field strength, propagation length and the Verdet constant. However, Fig. 3 reveals the twisting-induced transmissivity also plays a critical role. The ratio of the minor axis to the major axis of elliptical polarization is proportional to $|(A_L - A_R)/(A_L + A_R)|$, where $A_L$ and $A_R$ represent the amplitude of the transmitted LCP and RCP, respectively. Decomposing the linearly polarized incident wave into the LCP and RCP, high transmissivity implies $A_L$ and $A_R$ are nearly equal. Thus, the ratio approaches zero, indicating a linearly polarized output EM wave.

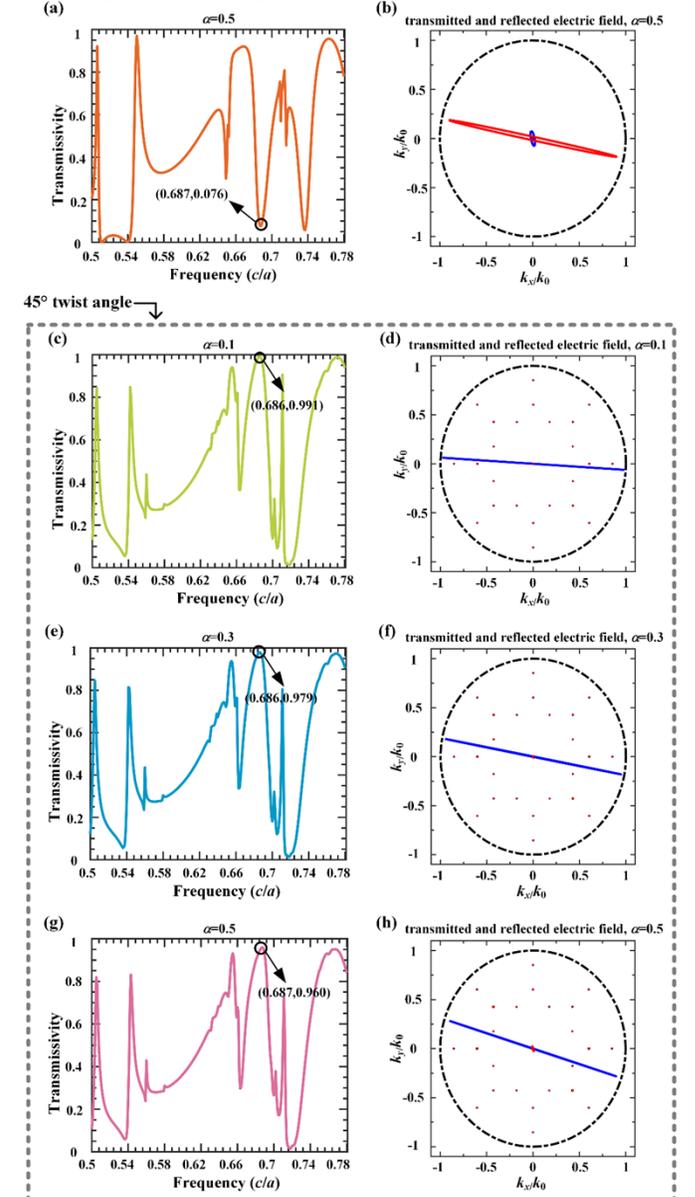

**Fig. 4.** (a, c, e, g) Transmission spectrum under different conditions: (a) $\alpha$=0.5, 0° twist angle; (c) $\alpha$=0.1, 45° twist angle; (e) $\alpha$=0.3, 45° twist angle; (g) $\alpha$=0.5, 45° twist angle. All diffractions of the transmitted (blue) and reflected (red) electric field within the light cone under the following conditions: (b) at frequency 0.687 $c/a$, $\alpha$=0.5, 0° twist angle; (d) at frequency 0.686 $c/a$, $\alpha$=0.1, 45° twist angle; (f) at frequency 0.686 $c/a$, $\alpha$=0.3, 45° twist angle; (h) at frequency 0.687 $c/a$, $\alpha$=0.5, 45° twist angle.

By combining the tuning effect of the twist angle on resonant frequencies, the twisted slabs can realize perfect linear polarization of the transmitted waves. When the incident EM waves are polarized along the *x*-axis, Fig.4 shows the transmission and diffraction for different $\alpha$ and twist angles. Figs. 4(a) and (g) show the transmission spectra of 0° and 45° twist angle with $\alpha$=0.5. Due to low transmissivity at 0.687 $c/a$ with 0° twist angle, the transmitted and reflected waves display the elliptical polarization state. At the same frequency, the 45° twist angle improves the transmissivity. Fig. 4(h) shows the zeroth-order transmitted wave is perfectly linear polarized and the high-order diffraction modes are suppressed.

Comparing Figs. 4(c), (e) and (g), $\alpha$ affects the peak values of the transmission spectra. In Fig. 4(c), the transmissivity reaches a maximum of 0.991 at 0.686 $c/a$. When $\alpha$ rises to 0.3 and 0.5, the maximum becomes 0.979 and 0.960, respectively. Due to the high transmissivity at the resonant peaks, the polarizations of the zeroth-order transmitted diffraction waves maintain a perfect linear polarization, as shown in Figs. 4(d), (f) and (h). And, the rotation angles are apparently different. As $\alpha$ increases from 0.1 to 0.3 and 0.5, the polarization of the zeroth-order transmitted wave is rotated by -3.481°, -10.785° and -17.448° relative to the incident wave. The rotation angle in Faraday rotation is linearly dependent on the external magnetic field. For each increase of 0.2 in $\alpha$, the polarization of the zeroth-order transmitted wave rotates by an additional 7° approximately. Moreover, the transmitted energy is predominantly concentrated in the zeroth-order diffraction, and the high-order diffractions are too weak to distinguish their polarizations.

In conclusion, TBMPC demonstrates a unique synergy between twist engineering and magneto-optical effects, enabling unprecedented control over light-matter interactions. By applying an external magnetic field and systematically varying the twist angle, the platform achieves distinct circular dichroism and tunable Faraday rotation. Significant chiral characteristics are distinctly observed at the resonance frequencies, which can be modulated by the twist angle. The output zeroth-order diffraction waves with linear polarization tend to occur under conditions of high transmissivity. By tailoring the transmissivity via the twist angle, the transmitted wave achieves perfect linear polarization. At the 45° twist angle, the high-order diffractions at resonant frequencies are suppressed, avoiding crosstalk to the zeroth-order polarization. The ability of the study system to modulate transmission characteristics with high degrees of freedom highlights its potential for reconfigurable optical devices and inspires innovations in non-reciprocal components and polarization-selective sensors. The geometric and magnetic tuning can pave the way for advanced topological photonic systems.

**Funding.** This work is supported by the NSFC with Grant No. 12322412.

**Disclosures.** The authors declare no conflicts of interest.

**Data availability.** Data of the present work could be obtained from the authors upon reasonable request.